# Surface electromagnetic waves near a black hole event horizon and their observational consequences

**Igor I. Smolyaninov**

ECE Department, University of Maryland, College Park MD 20740, USA; smoly@umd.edu

**Abstract:** Localization phenomena in light scattering from random fluctuations of matter fields and space–time metric near a black hole horizon were predicted to produce a pronounced peak in the angular distribution of second harmonic light in the direction normal to the horizon. Therefore, detection of second harmonic generation may become a viable observational tool to study spacetime physics near event horizons of astronomical black holes. The light localization phenomena near the horizon may be facilitated by the existence of surface electromagnetic wave solutions. In this communication we study such surface electromagnetic wave solutions near the horizon of a Schwarzschild metric describing a black hole in vacuum. We demonstrate that such surface wave solutions must appear when quantum gravity effects are taken into account. Potential observational evidence of this effect are also discussed.

## 1. Introduction

During the recent several years an enormous progress has been made in imaging and exploration of spatial regions located very near the event horizons of known astronomical black holes [1-3]. However, many more experimental tools will be required in the future to observe and study the complicated physics of these highly non-trivial spacetime regions in more detail. In this communication we will concentrate on further development of a recent proposal [4] to use optical second harmonic (SH) generation as an alternative observational tool to explore gravitational physics in the immediate vicinity of an event horizon of a black hole. In ref. [4] it was demonstrated that SH generation must be strongly enhanced when light from a distant star interacts with random fluctuations of various fields (such as matter fields and fluctuating spacetime metric) near an event horizon. This strongly directional SH emission occurs due to localization phenomena in light scattering from these fluctuations. It is directed away from the black hole perpendicular to the horizon, and therefore it has very good chances of escaping from the black hole. As illustrated in Fig.1, this effect resembles the enhancement of SH radiation from randomly rough metal surfaces, which also occurs in the normal direction to a metal surface [5,6]. The SH peak in the normal direction occurs under external spatially coherent irradiation at any illumination direction, provided that the randomly rough surface is capable of supporting surface electromagnetic waves. When an external light source of frequency $\omega$ illuminates such a random surface, it couples into a system of weakly localized surface electromagnetic modes. Let us consider one of these surface modes which has a momentum $k$ along the interface. Upon surface propagation, this mode will experience a lot of backscattering, leading to generation of counterpropagating surface modes with momentum approximately equal to $-k$. When these counterpropagating modes of frequency $\omega$ interact via any kind of surface optical nonlinearity, they generate $2\omega$ light. Due to momentum conservation, this SH light may only have nonzero wave vector component in the direction perpendicular to the surface, leading to observation of the strongly enhanced diffuse SH generation in this direction. According to experimental observations [6], the angular width of this SH peak can be as small as a few degrees with respect to the mean normal, and its intensity appears to be march larger compared to the diffuse omnidirectional SH background. While potential observability of this very interesting effect has been established in [4], the theoretical consideration in that work was limited only to the case of Rindler metric, which describes the immediate vicinity of the black hole event horizon. In this communication we perform a more realistic study of surface electromagnetic wave solutions near the horizon of a Schwarzschild metric describing a "real" astronomical black hole in vacuum. We demonstrate that surface wave solutions must also appear under these more realistic conditions (especially when quantum gravity effects are taken into account). Potential observational evidence of this effect will be also discussed.

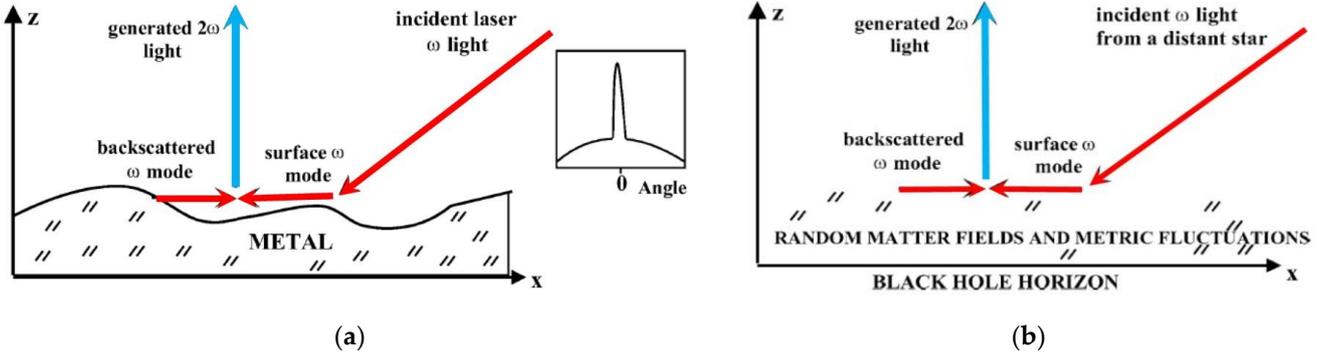

**Figure 1.** (**a**) Randomly rough metal surfaces exhibit strongly enhanced SH generation in the direction perpendicular to the surface under illumination with spatially coherent fundamental light. The inset illustrates typical distribution of the diffuse SH generation with respect to the normal angle to the metal surface. (**b**) In a similar fashion, SH generation must be strongly enhanced when light from a distant star interacts with random fluctuations of matter fields and spacetime metric near the event horizon. Escape of the SH light from the black hole is facilitated by the strongly directional character of this effect [4].

## 2. Results

In order to demonstrate existence of surface electromagnetic wave solutions near a black hole event horizon we must solve Maxwell equations in vacuum in the presence of gravitational field [7]:

$$\frac{\partial F_{ik}}{\partial x^l} + \frac{\partial F_{li}}{\partial x^k} + \frac{\partial F_{kl}}{\partial x^i} = 0 \quad \text{and} \quad (1)$$

$$\frac{1}{\sqrt{-g}}\frac{\partial}{\partial x^k}\left(\sqrt{-g}F^{ik}\right) = 0 \ , \quad (2)$$

where $F_{ik}$ is the electromagnetic field tensor.

Consideration of electromagnetic wave propagation near a black hole event horizon in [4] was based on the well-known analogy between these Maxwell equations in a curvilinear spacetime metric $g_{ik}(x, t)$ and the macroscopic Maxwell equations describing electromagnetic fields in the presence of matter background with some non-trivial electric permittivity tensor $\varepsilon_{ij}(x, t)$ and magnetic permeability tensors $\mu_{ij}(x, t)$ [7]. For example, the equations of electrodynamics in the presence of static gravitational field look exactly like Maxwell equations in some macroscopic electrodynamic medium in which

$$\varepsilon = \mu = g_{00}^{-1/2} \ , \quad (3)$$

where $g_{00}$ is the temporal component of the metric tensor. Following this approach, let us consider the static Schwarzschild metric describing an astronomical black hole in vacuum:

$$ds^2 = \left(1 - \frac{r_s}{r}\right)dt^2 - \frac{dr^2}{\left(1 - \frac{r_s}{r}\right)} - r^2\left(d\theta^2 + \sin^2\theta d\phi^2\right) \ , \quad (4)$$

where $r_s = 2\gamma M/c^2$ is the Schwarzschild radius of the black hole [7]. The corresponding equivalent material parameters are

$$\varepsilon = \mu = \frac{1}{\sqrt{1-\frac{r_s}{r}}} \tag{5}$$

Let us consider solutions of the macroscopic Maxwell equations in such a spherically symmetric geometry. The spatial variables in the Maxwell equations written in the spherical coordinates partially separate, and without the loss of generality we may assume field dependencies proportional to $e^{i(m\phi-\omega t)}$, where $m$ is integer. The macroscopic Maxwell equations may be written using the spherical coordinates $(r,\theta,\phi)$ as [8]:

$$\frac{1}{r\sin\theta}\left[\frac{\partial}{\partial\theta}(\sin\theta E_\phi) - imE_\theta\right] = \frac{i\omega\mu}{c}H_r \tag{6}$$

$$\frac{im}{r\sin\theta}E_r - \frac{1}{r}\frac{\partial}{\partial r}(rE_\phi) = \frac{i\omega\mu}{c}H_\theta \tag{7}$$

$$\frac{1}{r}\frac{\partial}{\partial r}(rE_\theta) - \frac{1}{r}\frac{\partial E_r}{\partial\theta} = \frac{i\omega\mu}{c}H_\phi \tag{8}$$

$$\frac{1}{r\sin\theta}\left[\frac{\partial}{\partial\theta}(\sin\theta H_\phi) - imH_\theta\right] = -\frac{i\omega\varepsilon}{c}E_r \tag{9}$$

$$\frac{im}{r\sin\theta}H_r - \frac{1}{r}\frac{\partial}{\partial r}(rH_\phi) = -\frac{i\omega\varepsilon}{c}E_\theta \tag{10}$$

$$\frac{1}{r}\frac{\partial}{\partial r}(rH_\theta) - \frac{1}{r}\frac{\partial H_r}{\partial\theta} = -\frac{i\omega\varepsilon}{c}E_\phi \tag{11}$$

$$\frac{1}{r}\frac{\partial}{\partial r}(r^2\varepsilon E_r) + \frac{\varepsilon}{\sin\theta}\frac{\partial}{\partial\theta}(\sin\theta E_\theta) + \frac{im\varepsilon}{\sin\theta}E_\phi = 0 \tag{12}$$

$$\frac{1}{r}\frac{\partial}{\partial r}(r^2\mu H_r) + \frac{\mu}{\sin\theta}\frac{\partial}{\partial\theta}(\sin\theta H_\theta) + \frac{im\mu}{\sin\theta}H_\phi = 0 \tag{13}$$

These equations may be simplified if we assume that $m=0$, which may be achieved for any given light ray by choosing the proper system of coordinates. Moreover, if $m=0$ the TM and TE polarized solutions may be separated, so that for the TM solutions (for which $E_\phi=H_r=H_\theta=0$) we obtain:

$$E_r = \frac{ic}{\omega\varepsilon r\sin\theta}\frac{\partial}{\partial\theta}(\sin\theta H_\phi) \quad, \text{and} \tag{14}$$

$$E_\theta = -\frac{ic}{\omega\varepsilon r}\frac{\partial}{\partial r}(rH_\phi) \tag{15}$$

Substitution of Eqs.(14,15) into Eq.(8) gives rise to the following wave equation for the TM polarized light:

$$-\frac{\varepsilon}{r}\frac{\partial}{\partial r}\left(\frac{1}{\varepsilon}\frac{\partial}{\partial r}(rH_\phi)\right) - \frac{1}{r^2}\frac{\partial}{\partial\theta}\left(\frac{1}{\sin\theta}\frac{\partial}{\partial\theta}(\sin\theta H_\phi)\right) = \frac{\varepsilon\mu\omega^2}{c^2}H_\phi \tag{16}$$

Let us search for approximate solutions of Eq.(16) which have the following functional form:

$$H_\phi \sim e^{ik_\theta r\theta} \quad, \tag{17}$$

where $rk_\theta \gg 1$ is assumed. Under such an assumption the wave equation may be re-written as

$$-\frac{\varepsilon}{r}\frac{\partial}{\partial r}\left(\frac{1}{\varepsilon}\frac{\partial}{\partial r}(rH_\phi)\right) - \frac{\varepsilon\mu\omega^2}{c^2}H_\phi = -k_\theta^2 H_\phi \tag{18}$$

which may be recast as a one-dimensional Schrödinger equation for an effective wave function defined as $\psi = H_\phi r/\sqrt{\varepsilon}$:

$$-\frac{\partial^2\psi}{\partial r^2} + \left(-\frac{1}{2\varepsilon}\frac{\partial^2\varepsilon}{\partial r^2} + \frac{3}{4\varepsilon^2}\left(\frac{\partial\varepsilon}{\partial r}\right)^2 - \varepsilon\mu\frac{\omega^2}{c^2}\right)\psi = -\frac{\partial^2\psi}{\partial r^2} + V\psi = -k_\theta^2\psi \quad, \tag{19}$$

In the latter equation $V$ plays the role of an effective potential and $-k_\theta^2$ plays the role of a total energy. Based on the expression for $\varepsilon$ and $\mu$ from Eq.(5), the effective potential near a Schwarzschild black hole equals

$$V = -\frac{\omega^2}{c^2}\frac{1}{\left(1-\frac{r_s}{r}\right)} - \frac{r_s\left(1-\frac{r_s}{4r}\right)}{2r^3\left(1-\frac{r_s}{r}\right)^2} + \frac{3r_s^2}{16r^4\left(1-\frac{r_s}{r}\right)^2} \approx -\frac{\omega^2 r_s}{c^2\rho} - \frac{3}{16\rho^2} \quad , \qquad (20)$$

where we have assumed that $r=r_s+\rho$ and $\rho \ll r_s$. This potential (plotted schematically in Fig.2) appears to be real and well-behaved even below the horizon, even though the effective $\varepsilon$ and $\mu$ parameters themselves are imaginary (as defined by Eq.(5)). Moreover, as noted in [4], the divergence of this potential at $\rho=0$ is supposed to be tamed by the quantum gravity effects. Note that such a situation is not unusual in macroscopic electrodynamics. A similar situation is observed in recently discovered surface electromagnetic waves guided by strongly lossy gradient structures [9]. We should also mention that the spacetime metric of a black hole interior is typically obtained as an analytical continuation of the conventional Schwarzschild metric (Eq.(4)) above the horizon, in which the dynamical roles of the temporal and radial coordinates are interchanged [10]. It is easy to verify that such an exchange does not affect Eqs.(1,2). Therefore, the effective potential described by Eq.(20) appears to be reasonably well justified.

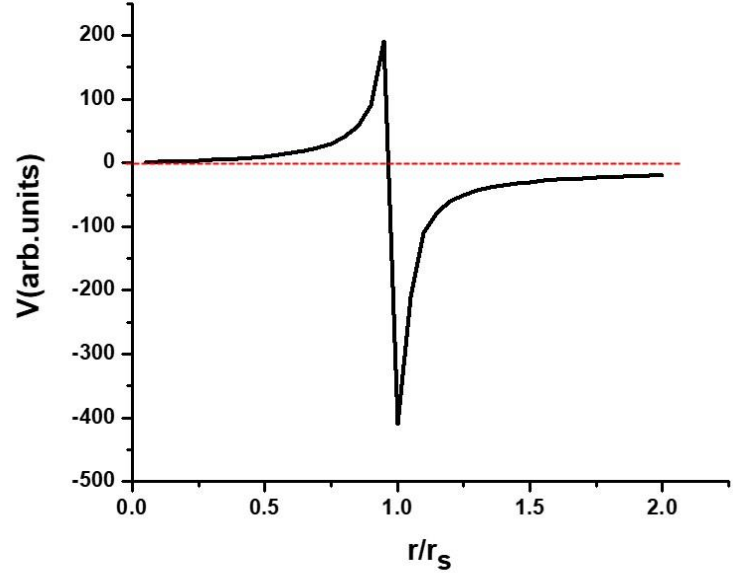

**Figure 2.** Basic shape of the effective potential from Eq.(20) near the horizon.

Let us analyze the eigenstate solutions of this effective Schrodinger equation. We should note that very near the horizon the potential energy term proportional to $1/\rho^2$ will always dominate. However, this only happens at very short distances when

$$\rho \ll \frac{3\lambda^2}{64\pi^2 r_s} \quad , \qquad (21)$$

where $\lambda$ is the wavelength of light at infinity. At larger distances the effective potential is Coulomb-like. An extensive analysis of solutions of the Schrödinger equation with a "Coulomb plus inverse-square potential"

$$V = -\frac{B}{\rho} + \frac{A}{\rho^2} \qquad (22)$$

may be found in [11]. If $A > -1/8$, the energy eigenstates of this potential are

$$E(n) = -\frac{2B^2}{\left(2n-1+\sqrt{1+8A}\right)^2} \qquad (23)$$

In contrast, if $A < -1/8$, the $\rho$-component of the spatial frequency diverges near $\rho=0$, and there is no finite ground state (the particle falls into the horizon). Since in our case $A = -3/16 < -1/8$ (see Eq.(20)), we have obtained a familiar conclusion that there are no zero-geodesics near the surface of a black hole within the scope of classical treatment of the Schwarzschild metric. In the absence of quantum gravity effects, every photon falls towards the horizon, and it is inevitably absorbed by the black hole. However, based on the general properties of 1D Schrödinger equations [12], it is well known that the case of $1/\rho^2$ potential is a borderline case, which separates potential wells exhibiting finite ground energy states from the much more divergent potential wells in which a finite ground state may not exist. In particular, any potential well that is weaker then $1/\rho^2$ exhibits a finite ground state. Therefore, emergence of a cutoff for any particular reason in the divergent $1/\rho^2$ behavior of the potential $V(\rho)$ near a black hole event horizon (for example, due to emergence of quantum gravitational minimum length $l_{min}$ of the order of the Planck scale) leads to emergence of a well-defined ground eigenstate among the wavefunctions described by Eq. (19). This eigenstate located at $k_\theta \sim l_{min}^{-1}$ gives rise to a fundamental guided surface electromagnetic wave propagating near the horizon. Note that the field configuration and the dispersion law of this mode strongly resembles the charge density wave in a gradient waveguide described in [9]. In addition, a set of well-defined excited surface states will also appear in this limit, which at large $n$ will tend to

$$E(n) = -k_{\theta n}^2 \approx -\frac{B^2}{2n^2} \qquad (24)$$

due to the Coulomb-like character of the attractive potential $V(\rho)$ at large $\rho$. The dispersion law of these excited modes appears to be

$$k_{\theta n} \approx \frac{\omega^2 r_s}{\sqrt{2} c^2 n} \qquad (25)$$

Thus, our detailed consideration reveals a family of surface electromagnetic wave modes near the horizon which justifies the proposal [4] to use optical SH generation (mediated by these modes) as an observational tool to explore gravitational physics in the immediate vicinity of the horizon.

## 3. Discussion

The presence of surface electromagnetic wave solutions near the horizon in the more general framework of a Schwarzschild metric further justifies the theory of SH generation near horizon, which was developed in [4]. This theory is virtually identical to the theory of strongly enhanced directional SH generation from randomly rough metal surfaces [5]. The latter theory was confirmed in the experiments with randomly rough metal surfaces capable of supporting surface electromagnetic waves [6], in which a narrow SH peak in the normal direction was observed under external spatially coherent laser illumination at any illumination direction. This interesting effect appears to be a generic property of systems which support surface electromagnetic modes and exhibit weak disorder. An essential similarity between the surface plasmon geometry described in [5,6] and an astronomical situation in which light from a distant star interacts with a black hole horizon (see Fig.1) arises from the fact that the illuminating light in both cases has a very high degree of spatial coherence. Since photon interference is the root cause of localization effects in light scattering, such effects are only possible for spatially coherent illumination. Distant stars indeed provide a source of such spatially coherent illumination (which is somewhat similar to laser light) because of their small angular dimensions.

It is also important that SH light directed perpendicular to the event horizon has the most chances to leave the neighborhood of a black hole. As a result, such a SH light may become a dominant component of its visible emission. Indeed, the omnidirectionally scattered fundamental ($\omega$) and SH ($2\omega$) light will be predominantly reabsorbed by the black hole. As a result, for a distant observer the intensity of the diffuse fundamental and SH light will be substantially attenuated in comparison with the directional second harmonic peak. While observation of such SH emission may not be easy, the described effect may be used to obtain unique experimental information on the inner workings of quantum gravity. Similar to Hawking radiation, the described SH emission is caused by the quantum gravitational effects at very high spatial frequencies (of the order of the Planck scale). On the other hand, localization effects induced by the quantum spacetime fluctuations, which are necessary for the directional SH generation to occur, do not play a substantial role in the Hawking radiation. Another important distinction between these two effects is that Hawking radiation is an internal property of a black hole, while the directional SH radiation represents a black hole reaction on the external illumination. Therefore, the described directional SH generation and the Hawking radiation are two very different effects of quantum gravity.

Another potentially interesting consequence of the newly obtained surface electromagnetic wave solutions, which look very similar to the charge density waves in gradient waveguides [9], is that we may potentially assign an interesting direct physical meaning to the holographic principle [13]. According to this guiding principle of quantum gravity, the physical description of a volume of space can be thought of as encoded on a surface bounding this volume, such as a gravitational horizon. As we have seen from Eq.(5), from the electromagnetic point of view a horizon corresponds to a surface where $\varepsilon=\mu\rightarrow\infty$, which means that a horizon must act as an electric and magnetic mirror (see Fig.3). For example, an electric charge $q$ located at a distance $l$ from the horizon must induce a redistribution of surface charge density $\sigma$ on the horizon defined as

$$\sigma(x,y) = -\frac{ql}{2\pi\left(x^2+y^2+l^2\right)^{3/2}}, \tag{26}$$

leading to the appearance of "image charge" in the mirror. As a result, the real volume physics will be dutifully "reflected" by 2D redistribution of these surface charges. Excitation and propagation of the newly obtained surface charge density waves at the horizon may provide a real physical mechanism behind the holographic imaging. Note that a self-force acting on an accelerated charge indeed has a contribution, which may be interpreted as a Coulomb force from an image charge reflected in a distant Rindler horizon (see for example [14]).

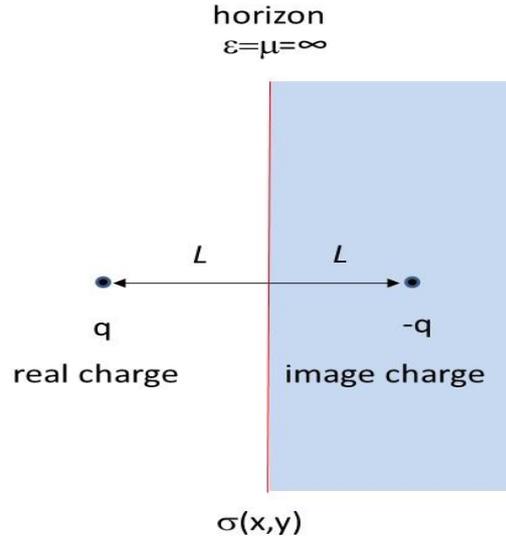

**Figure 3.** Appearance of image charges reflected in the mirror is caused by propagation and redistribution of surface charges $\sigma$.

As far as observational consequences of the black hole SH emission are concerned, it may potentially be detected as a weak anomalously blue-shifted light source which total brightness must be proportional to the surface area of the black hole and the square of the total incident light intensity provided by nearby light sources. Assuming that the optical nonlinearities near horizon must be extremely strong due to enormous field intensities near the horizon, the second harmonic conversion efficiency may be considerable (for example, in some solid-state physics situations SH conversion efficiency reaches up to ~10%). This argument provides reasonably optimistic chances of detection of SH light generated by a black hole located in a dense stellar association or in a center of a bright galaxy.

We should also mention that experimental observations of SH light from various astronomical sources are not unusual. In particular, we can mention observations of SH of the cyclotron line in the spectra of such high-mass bright X-ray sources as 4U1907+09, which were obtained by the BeppoSAX [15]. Since such highly massive bright X-ray sources are widely believed to be powered by black holes, these and similar experimental observations may potentially need to be re-analyzed in search for the evidence of the highly interesting quantum gravity effects described above.

Even though it may be difficult to observe these effects in astronomical observations, one might also observe these effects in the higher-dimensional mini-black holes on the Planck mass scale, which would be observed in future accelerator experiments [16]. When the black hole mass approaches the order of the Planck mass due to the Hawking radiation, it would be expected that quantum gravity effects would also lead to quantum fluctuations in the background metric. In four dimensions, such modified background geometries would be given by the quantum deformed Schwarzschild black holes [17], the black holes in the noncommutative models [18], and black holes in the asymptotically safe gravity [19].

## 4. Conclusions

In conclusion, we have demonstrated that localization phenomena in light scattering from random fluctuations of matter fields and quantum spacetime in the vicinity of a black hole horizon may produce an intense peak of second harmonic light directed perpendicular and away from the horizon. Therefore, detection of second harmonic generation may become a viable observational tool to study spacetime physics near event horizons of an astronomical black hole. The light localization phenomena near the horizon may be facilitated by the existence of surface electromagnetic wave solutions. Such surface wave solutions must appear when quantum gravity effects are taken into account. Potential observational evidence of this effect has been discussed, which indicate experimental viability of the proposed technique.